\documentclass [11pt] {article}
\usepackage[dvips]{graphicx,color}
\usepackage[hyperindex,pdfmark,letterpaper]{hyperref}
\newcommand{\indentspace}{\phantom{aaa}}
\newcommand{\lchoose}[2]{{#1\choose #2}}
\newcommand{\bfc}{{\bf c}}

\newcommand{\bfx}{{\bf x}}
\newcommand{\bfy}{{\bf y}}
\newcommand{\bfone}{{\bf 1}}

\newcommand{\cale}{{\cal E}}
\newcommand{\calj}{{\cal J}}
\newcommand{\calk}{{\cal K}}
\newcommand{\call}{{\cal L}}

\newcommand{\smallfrac}[2]{\frac{\mbox{\footnotesize $#1$}}{\mbox{\footnotesize $#2$}}}

\definecolor{Red}{rgb}{1,0,0}
\definecolor{Blue}{rgb}{0,0,1}

\usepackage{verbatim}

\begin{document}
\title{
  \begin{flushleft}
    {\footnotesize BU-CCS-991001}\\[0.5cm]
  \end{flushleft}
  \bf A Particulate Basis for an Immiscible Lattice-Gas Model}
\author{
  Bruce M. Boghosian\\
  {\footnotesize Center for Computational Science,}\\
  {\footnotesize Boston University, 3 Cummington Street,
    Boston, Massachusetts 02215, U.S.A.}\\
  {\footnotesize{\tt bruceb@bu.edu}}\\[0.3cm]
  Peter V. Coveney\\
  {\footnotesize Centre for Computational Science,}\\
  {\footnotesize Queen Mary and Westfield College,
    University of London,}\\
  {\footnotesize Mile End Road, London E1 4NS, U.K.}\\
  {\footnotesize{\tt p.v.coveney@qmw.ac.uk}}\\
  }

\date{\today}
\maketitle

\begin{abstract}
  We show that a phenomenological hydrodynamic lattice-gas model of
  two-phase flow, developed by Rothman and Keller in 1988 and used
  extensively for numerical simulations since then, can be derived from
  an underlying model of particle interactions.  From this result, we
  elucidate the nature of the hydrodynamic limit of the Rothman-Keller
  model.
\end{abstract}

\vspace{0.2truein}

\vspace{0.3in}
\section{Introduction}

In 1986 it was discovered that certain mass and momentum conserving
lattice-gas automata gave rise to the isotropic Navier-Stokes equations
in the hydrodynamic limit~\cite{bib:fhp,bib:swolf}.  In 1988, Rothman
and Keller extended this discovery by introducing a hydrodynamic
lattice-gas model of immiscible fluids~\cite{bib:rk}.  Their model, and
lattice Boltzmann variants thereof, have become an important tool for
simulating the hydrodynamics of multiphase flow~\cite{bib:rz}.  In both
the original and RK lattice-gas models, the dynamics can be decomposed
into two steps: In the first, the particles {\it propagate} along the
lattice vectors to new sites; in the second, the particles entering each
site {\it collide} by redistributing mass and momentum.

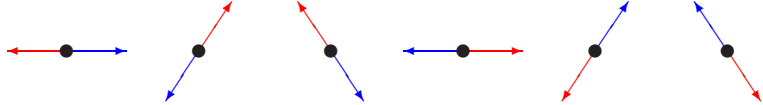
\begin{figure}
\centering
\begin{picture}(345,50)(0,0)
\put(45.,18.){
\begin{picture}(36,36)(0,0)
\put(0.,0.){\color{Red}\vector(-1,0){22.5}}
\put(0.,0.){\color{Blue}\vector(1,0){22.5}}
\put(0.,0.){\circle*{5.}}
\end{picture}}
\put(95.,18.){
\begin{picture}(36,36)(0,0)
\put(0.,0.){\color{Red}\vector(2,3){12.5}}
\put(0.,0.){\color{Blue}\vector(-2,-3){12.5}}
\put(0.,0.){\circle*{5.}}
\end{picture}}
\put(145.,18.){
\begin{picture}(36,36)(0,0)
\put(0.,0.){\color{Red}\vector(-2,3){12.5}}
\put(0.,0.){\color{Blue}\vector(2,-3){12.5}}
\put(0.,0.){\circle*{5.}}
\end{picture}}
\put(195.,18.){
\begin{picture}(36,36)(0,0)
\put(0.,0.){\color{Blue}\vector(-1,0){22.5}}
\put(0.,0.){\color{Red}\vector(1,0){22.5}}
\put(0.,0.){\circle*{5.}}
\end{picture}}
\put(245.,18.){
\begin{picture}(36,36)(0,0)
\put(0.,0.){\color{Blue}\vector(2,3){12.5}}
\put(0.,0.){\color{Red}\vector(-2,-3){12.5}}
\put(0.,0.){\circle*{5.}}
\end{picture}}
\put(295.,18.){
\begin{picture}(36,36)(0,0)
\put(0.,0.){\color{Blue}\vector(-2,3){12.5}}
\put(0.,0.){\color{Red}\vector(2,-3){12.5}}
\put(0.,0.){\circle*{5.}}
\end{picture}}
\end{picture}
\caption{{\bf One Set of Possible Collision Outcomes for Immiscible
    Fluid LGA:} The colors denote the particles' ``color charge.''  The
  incoming state had one particle of each color and zero net momentum.
  Note that there are six possible outcomes that conserve these
  quantities.}
\label{fig:colls}
\end{figure}

In the Rothman-Keller (RK) model, the masses of the various immiscible
fluid species and the total momentum are conserved locally, but the
choice of collision outcome at each site of the RK model depends on the
water-minus-oil order parameter, or ``color,'' of the neighboring sites.
See Fig.~\ref{fig:colls} for one set of possible collision outcomes that
might be allowed by the conservation laws.  The {\it flux} of the color
is determined for each such outgoing state, and its local gradient or
{\it field} is determined by examining the neighboring sites.  The
negative of the dot product of this color flux and color field is then a
measure of the propensity of outgoing particles to move to sites
dominated by particles of their own type, and was called the {\it color
  work} by RK; their prescription was then to choose the outcome that
minimizes this work in order to create cohesion and interfacial tension.
(In case of a tie, the outcome is chosen randomly from among the states
with minimal color work.)  In later work, Chen, Chen, Doolen and
Lee~\cite{bib:chen}, and Chan and Liang~\cite{bib:cl} noted that this
minimization of color work is really just the low-temperature limit of a
Boltzmann sampling procedure.

This paper proposes a microscopic interpretation of the RK model.  In
the limit where the ratio of the mean-free path to the interaction range
is small, we show that an arbitrary interaction potential can be
expanded in terms of dot products of fluxes and fields, of which the RK
model is merely the first term.  This observation accounts for much of
the success and utility of the RK model in describing the hydrodynamics
of multiphase flow.

\section{Hydrodynamic Lattice-Gas Automata}
\label{sec:hlga}

In lattice-gas models of hydrodynamics, incoming particles {\it collide}
at each site $\bfx$ of a lattice $\call$, in a manner to be discussed at
length shortly, and then {\it propagate} to one of $B$ neighboring sites
$\bfx+\bfc_i\in\call$, where $i\in\{1,\ldots,B\}$.  (Note that some of
the $\bfc_i$'s may be zero in order to accommodate ``rest particles'' in
the model.)  We suppose that the occupancy of each of these $B$ channels
can be represented by $L$ bits $n_i^\ell (\bfx)\in\{ 0,1\}$, where
$i\in\{1,\ldots,B\}$ and $\ell\in\{1,\ldots,L\}$.  Throughout the
remainder of this paper, we shall illustrate various concepts by
applying them to three concrete examples:
\begin{itemize}
\item {\bf Example 1:} In a lattice gas for a single-species
  Navier-Stokes fluid~\cite{bib:fhp,bib:swolf,bib:fchc}, we take one bit
  ($L=1$) in each direction that represents the presence or absence of a
  particle moving in that direction.
\item {\bf Example 2:} In a lattice gas for two immiscible
  fluids~\cite{bib:rk}, on the other hand, we might take two bits per
  direction ($L=2$) so that there can be one bit for water particles
  $n_i^W(\bfx)$ and one bit for oil particles $n_i^O(\bfx)$ in each
  direction.
\item {\bf Example 3:} We consider the model of Chen, Chen, Doolen and
  Lee~\cite{bib:chen}, in which there is one bit in each direction
  ($L=1$), one ``rest'' particle direction $i=R$ such that $\bfc_R=0$,
  and only the rest particles feel an interaction potential.
\end{itemize}

We suppose that there is a charge-like attribute $q_i(\bfx)$ associated
with the bits in direction $i$ at site $\bfx$, and we specialize to
potential energies of the form
\begin{equation}
 V = \smallfrac{1}{2}\sum_{\bfx,\bfy}\sum_{i,j}
     q_i(\bfx)q_j(\bfx+\bfy)\phi(|\bfy|),
 \label{eq:pot}
\end{equation}
where the factor of $1/2$ prevents double counting.  In Example 1, the
charge-like attribute might be equal to the channel occupancy
$n_i(\bfx)$.  In Example 2, on the other hand, the charge-like attribute
might be the order parameter
\begin{equation}
q_i(\bfx)=n_i^W(\bfx)-n_i^O(\bfx)
\end{equation}
which measures the excess of water over oil.  In Example 3, we take
$q_i(\bfx) = n_i(\bfx)\delta_{iR}$, since only the rest particles have a
charge-like attribute.  In what follows, we shall also have occasion to
refer to the {\it total} (summed over directions) charge-like attribute
of a site $q(\bfx)\equiv\sum_i q_i(\bfx)$.

As noted above, any fluid model has some set of quantities that must be
conserved in collisions.  In Examples 1 and 3, we should demand
conservation of the mass $\sum_i n_i(\bfx)$ and the momentum $\sum_i
n_i(\bfx) \bfc_i$.  In Example 2, on the other hand, collisions must
conserve water mass $\sum_i n^W_i(\bfx)$, oil mass $\sum_i n^O_i(\bfx)$
and total momentum $\sum_i [n^W_i(\bfx)+n^O_i(\bfx)] \bfc_i$.
Alternatively, we can say that all three examples conserve total mass
and the total momentum, while Example 2 also conserves the charge-like
attribute, $q(\bfx)$.  These conserved quantities naturally partition
the set of all states of a given site(s) into equivalence classes of
states with the same values for all of the conserved quantities.  For
example, Fig.~\ref{fig:colls}, relevant to Example 2 above, illustrates
the equivalence class comprised of the six possible collisional outcomes
that may result when one water particle and one oil particle enter a
single site on a two-dimensional triangular lattice from opposite
directions (and, hence, with zero total momentum).

\section{Collisional Energetics}
\label{sec:ce}

\begin{figure}
\centering
%\begin{picture}(210,530)(50,0)
\begin{picture}(210,225)(50,25)
% Put dots on lattice sites:
\put(100,60){\circle*{5}}
\put(60,120){\circle*{5}}
\put(180,180){\circle*{5}}
\put(260,180){\circle*{5}}
% Put thick arrows between dots:
\put(100,60){\thicklines\vector(4,3){160}}
\put(100,60){\thicklines\vector(-2,3){40}}
\put(60,120){\thicklines\vector(2,1){120}}
\put(260,180){\thicklines\vector(-1,0){80}}
% Put thin lattice lines about each dot:
\put(100,60){\thinlines\line( 1, 0){20}}
\put(100,60){\thinlines\line( 2, 3){10}}
\put(100,60){\thinlines\line(-2, 3){10}}
\put(100,60){\thinlines\line(-1, 0){20}}
\put(100,60){\thinlines\line(-2,-3){10}}
\put(100,60){\thinlines\line( 2,-3){10}}
\put(60,120){\thinlines\line( 1, 0){20}}
\put(60,120){\thinlines\line( 2, 3){10}}
\put(60,120){\thinlines\line(-2, 3){10}}
\put(60,120){\thinlines\line(-1, 0){20}}
\put(60,120){\thinlines\line(-2,-3){10}}
\put(60,120){\thinlines\line( 2,-3){10}}
\put(180,180){\thinlines\line( 1, 0){20}}
\put(180,180){\thinlines\line( 2, 3){10}}
\put(180,180){\thinlines\line(-2, 3){10}}
\put(180,180){\thinlines\line(-1, 0){20}}
\put(180,180){\thinlines\line(-2,-3){10}}
\put(180,180){\thinlines\line( 2,-3){10}}
\put(260,180){\thinlines\line( 1, 0){20}}
\put(260,180){\thinlines\line( 2, 3){10}}
\put(260,180){\thinlines\line(-2, 3){10}}
\put(260,180){\thinlines\line(-1, 0){20}}
\put(260,180){\thinlines\line(-2,-3){10}}
\put(260,180){\thinlines\line( 2,-3){10}}
% Put in site labels:
\put(98,70){$\bfx$}
\put(250,158){$\bfx+\bfy$}
\put(160,200){$\bfx+\bfy+\bfc_j$}
\put(50,140){$\bfx+\bfc_i$}
% Put in vector labels:
\put(175,123){$\bfy$}
\put(80,95){$\bfc_i$}
\put(215,172){$\bfc_j$}
\put(120,142){$\bfy+\bfc_j-\bfc_i$}
\end{picture}
\caption{{\bf Change in Potential Energy:} The change in the potential
  energy of interaction between two charges moving in (possibly)
  different directions at (possibly) different sites is illustrated
  here.  See Eq.~(\protect{\ref{eq:dv}}).}
\label{fig:dp}
\end{figure}
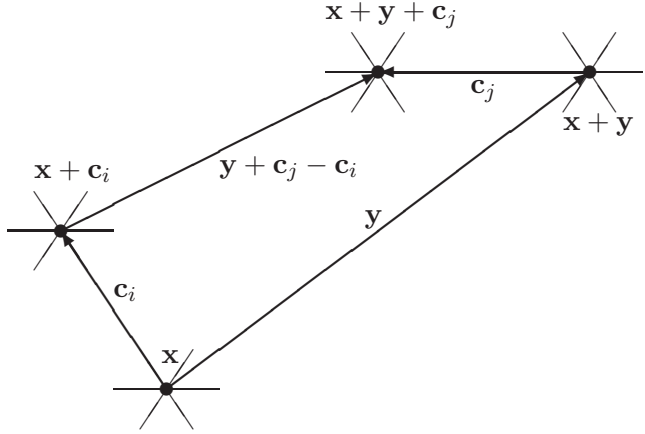

We denote the postcollision charge-like attribute with velocity $\bfc_i$
at site $\bfx$ by $q'_i(\bfx)$.  Upon subsequent propagation, the charge
$q'_i(\bfx)$ will be at position $\bfx+\bfc_i$, and the charge
$q'_j(\bfx+\bfy)$ will be at position $\bfx+\bfy+\bfc_j$.  This is
illustrated in Fig.~\ref{fig:dp}.  The change in the potential energy
due to {\it both} collision and propagation is then given by
\begin{eqnarray}
\Delta V &=&
\phantom{+}
\sum_{\bfx,\bfy}\sum_{i,j}
q'_i(\bfx)q'_j(\bfx+\bfy)\phi\left(\left|\bfy+\bfc_j-\bfc_i\right|\right)-
\sum_{\bfx,\bfy}\sum_{i,j}
q_i(\bfx)q_j(\bfx+\bfy)\phi\left(\left|\bfy\right|\right)\nonumber\\
&=&
\phantom{+}
\sum_{\bfx,\bfy}\sum_{i,j}
q'_i(\bfx)q'_j(\bfx+\bfy)
\left[\phi\left(\left|\bfy+\bfc_j-\bfc_i\right|\right)-
      \phi\left(\left|\bfy\right|\right)\right]\nonumber\\
& &
+\sum_{\bfx,\bfy}\sum_{i,j}
\left[q'_i(\bfx)q'_j(\bfx+\bfy)-q_i(\bfx)q_j(\bfx+\bfy)\right]
\phi\left(\left|\bfy\right|\right)\nonumber\\
&=&
\phantom{+}\Delta V_c + \Delta V_n,
\label{eq:dv}
\end{eqnarray}
where we have defined the contribution to $\Delta V$ due to the movement
of the interacting particles,
\begin{equation}
\Delta V_c\equiv
\sum_{\bfx,\bfy}\sum_{i,j}
q'_i(\bfx)q'_j(\bfx+\bfy)
\left[\phi\left(\left|\bfy+\bfc_j-\bfc_i\right|\right)-
      \phi\left(\left|\bfy\right|\right)\right],
\label{eq:dvc}
\end{equation}
and that due to nonconservation of the charge-like attribute,
\begin{equation}
\Delta V_n\equiv
\sum_{\bfx,\bfy}
\left[q'(\bfx)q'(\bfx+\bfy)-q(\bfx)q(\bfx+\bfy)\right]
\phi\left(\left|\bfy\right|\right).
\label{eq:dvn}
\end{equation}
Note that $\Delta V_c$ vanishes for systems in which the interacting
particles do not move, including our Example 3.  Likewise, note that
$\Delta V_n$ vanishes for systems with a conserved charge-like
attribute, including our Examples 1 and 2.

\section{The Flux-Field Decomposition}
\label{sec:ff}

If the potential kernel $\phi (y)$ is analytic, we may expand $\Delta
V_c$ in a Taylor series in the ratio of the characteristic lattice
spacing\footnote{Since lattice gases are usually dense fluids for which
  particles undergo a collision at every step, $c$ can also be thought
  of as a mean-free path.} $c$ to the characteristic interaction range
$y$.  This is done in Appendix~\ref{app:a} where it is shown that every
term of this series can be expressed as the complete inner product of a
tensor flux with a tensor field.  More specifically, we find that
\begin{equation}
\Delta V_c =
 \sum_\bfx
 \sum_{n=1}^\infty
 \sum_{r=\lceil n/2\rceil}^n
 \calj_r' (\bfx)
 \bigodot^r
 \cale_{n,r} (\bfx),
 \label{eq:dvcff}
\end{equation}
where we have defined the (local) $r$th-rank tensor flux of outgoing
particles,
\begin{equation}
\calj_r' (\bfx)
 \equiv
 \sum_i q_i' (\bfx)
 \left(
  \bigotimes^r\bfc_i
 \right),
 \label{eq:fluxes}
\end{equation}
and where the (generally nonlocal) $r$th-rank tensor fields are defined
in terms of $(n-r)$th-rank outgoing tensor fluxes of neighboring sites,
\begin{eqnarray}
\cale_{n,r}' (\bfx)
 & \equiv &
 \frac{(-1)^r}{(1+\delta_{n,2r})n!}\lchoose{n}{r}
 \sum_\bfy
 \calk_n (\bfy)
 \bigodot^{n-r}
 \calj_{n-r}' (\bfx+\bfy).
 \label{eq:fields}
\end{eqnarray}
In all the above expressions, primes are used to denote dependence on
postcollision values, $\bigotimes^r$ denotes an $r$-fold outer product,
$\bigodot^r$ denotes an $r$-fold inner product, and we have defined the
$n$th-rank completely symmetric kernel
\begin{equation}
\calk_n (\bfy)
 \equiv
 \sum_{m=\lceil n/2 \rceil}^n
 \frac{\phi_m(y)}{(n-m)!}\,
 \mbox{\rm per}
 \left[
  \left(\bigotimes^{2m-n}\bfy\right)
  \otimes
  \left(\bigotimes^{n-m}\bfone\right)
 \right],
\label{eq:kernel}
\end{equation}
where ``per'' indicates a summation over all distinct permutations of
indices, and where we have defined the following functions related to
the derivatives of $\phi (y)$:
\begin{equation}
\phi_m(y)\equiv
 \left(\frac{1}{y}\frac{d}{dy}\right)^m
 \phi(y).
\end{equation}
The first few such kernels are thus given by
\begin{eqnarray}
\left[\calk_1 (\bfy)\right]_{i}
 &=&
 \phi_1 (y) y_i
\\
\left[\calk_2 (\bfy)\right]_{ij}
 &=&
 \phi_1 (y) \delta_{ij} +
 \phi_2 (y) y_iy_j
\\
\left[\calk_3 (\bfy)\right]_{ijk}
 &=&
 \phi_2 (y) \left( y_i\delta_{jk} +
                   y_j\delta_{ik} +
                   y_k\delta_{ij}
            \right) +
 \phi_3 (y) y_iy_jy_k
\\
\left[\calk_4 (\bfy)\right]_{ijkl}
 &=&
 \phi_2 (y) \left( \delta_{ij}\delta_{kl} +
                   \delta_{ik}\delta_{jl} +
                   \delta_{il}\delta_{jk}
            \right)/2 +
\nonumber\\
& &
 \phi_3 (y) \left( y_iy_j\delta_{kl} +
                   y_iy_k\delta_{jl} +
                   y_iy_l\delta_{jk} +
                   y_jy_k\delta_{il} +
                   y_jy_l\delta_{ik} +
                   y_ky_l\delta_{ij}
            \right) +
\nonumber\\
& &
 \phi_4 (y) y_iy_jy_ky_l.
\end{eqnarray}

For completeness, we note that, since the zero-rank flux is just the
charge, the portion of $\Delta V$ arising from nonconservation of the
charge-like attribute may also be written in terms of these fluxes,
\begin{equation}
\Delta V_n =
 \smallfrac{1}{2}
 \sum_{\bfx,\bfy}
 \left[
  \calj_0'(\bfx)\calj_0'(\bfx+\bfy) -
  \calj_0(\bfx)\calj_0(\bfx+\bfy)
 \right]\phi_0(y)
\end{equation}
as follows immediately from Eqs.~(\ref{eq:dvn}) and (\ref{eq:fluxes}).

Because the fields themselves depend on the {\it post-collision} fluxes
at neighboring sites, it is generally necessary to include $\Delta V_n$
in the collisional energetics.  There are, however, some very useful
exceptions to this rule.  Some of the fluxes $\calj_r (\bfx)$ may be
conserved, in which case $\calj_r' (\bfx)=\calj_r (\bfx)$, since their
values for incoming and outgoing states must be identical.  If all of
the fluxes that go into the calculation of a field $\cale_{n,r}'(\bfx)$
are conserved quantities, then we can write $\cale_{n,r}' (\bfx) =
\cale_{n,r} (\bfx)$ as well.

For example, let us return to consider our Example 2.  At zeroth order,
Eq.~(\ref{eq:dvn}) indicates that $\Delta V_n$ vanishes because
$\calj_0$ is conserved.  At first order ($n=1$) the field
$\cale_{1,1}'(\bfx)$ depends only on the zero-order flux -- namely the
total order parameter (water minus oil) -- at a given site, and this is
conserved.  Hence, the first-order energy change can be computed for the
order-parameter flux for each outgoing state, and the field based on the
{\it incoming} states of the neighbors $\bfx+\bfy$.  If we restrict this
set of neighbors to immediately adjacent sites, and look no further than
lowest order in the Taylor expansion, we see that this reduces {\it
  precisely to the RK model}.  The RK model is thus an approximation
that is valid to first-order in $c/y$.  While this fact may explain much
of the utility of the RK model, in the following section we shall see
that it may also indicate that the scaling limit of the RK model is more
subtle than previously suspected.  In any case, this also means that an
alternative model that uses the exact potential energy, $\Delta V$ of
Eq.~(\ref{eq:dv}), to sample post-collision states at each site will
certainly be no worse than the RK model which is known to capture much
of the phenomenology of immiscible fluid dynamics.

Note that if {\it both} the flux and field in one of the terms of the
Taylor expansion can be evaluated based on incoming quantities, then
that term of $\Delta V$ is the same for all outgoing quantities, and
therefore does not discriminate between them, so that it is necessary to
go to higher order (at least until one encounters the first nonconserved
fluxes) to obtain any interaction at all.  We saw this with the
vanishing of $\Delta V_n$ for systems with conserved charge-like
attribute, such as Example 2 above.  To see this happen at higher order,
let us consider the simpler Example 1 which, perhaps for this very
reason, has been less studied.  Both fluxes $\calj_0$ and $\calj_1$ are
conserved, since they are the mass and momentum, respectively.  It
follows that the entire $n=1$ term of $\Delta V_c$ is the same for all
outgoing states.  Nontrivial interaction between particles therefore
does not even begin until the $n=2$ term of the expansion, since
$\calj_2$ (which is, in fact, related to the pressure tensor) is not a
conserved quantity.

\section{The Hydrodynamic Limit of the RK Model}

In order to obtain useful quantitative information from a hydrodynamic
lattice gas, one must be careful to work in the correct asymptotic
regime.  This usually involves scaling the various dimensionless
parameters of the problem with the Knudsen number
$\mbox{Kn}\sim\lambda/L$, where $\lambda$ is the mean-free path and $L$
is the characteristic size.  In incompressible Navier-Stokes flow, for
example, one desires that the Mach number $\mbox{M}\equiv U/C$, where
$U$ is the characteristic hydrodynamic velocity and $C$ is the sound
speed, scale with the Knudsen number $\mbox{M}\sim {\cal O}(\mbox{Kn})$.
Since the viscosity $\nu$ goes as the product of mean-free path and
sound speed, $\lambda C$, this implies that the Reynolds number
$\mbox{Re}$ scales as $\mbox{M}/\mbox{Kn}\sim {\cal O}(1)$\footnote{Note
  that this does not mean that the numerical value of the Reynolds
  number must be near unity.  Rather it means that Re approaches a
  constant value in the scaling limit.}.  We also demand that the
Strouhal number $\mbox{St}\equiv U\tau/L$ and the fractional density
fluctuation $\delta\rho/\rho_0$, where $\tau$ is the mean-free time and
$\rho_0$ is the average background density, both scale as ${\cal
  O}(\mbox{Kn}^2)$.  This limit is well known~\cite{bib:ll} to reduce
the compressible Navier-Stokes equations to their incompressible
counterparts.  Since, for a dense LGA, the mean-free path $\lambda$ goes
as the grid size $c$, this means that in order to approach the continuum
limit, every time one doubles the size of the lattice (halves
$\mbox{Kn}$), one must quadruple the number of time steps (since
$\mbox{St}$ is quartered), and verify that the fractional density
fluctuation (a measured ``output'' quantity in a lattice-gas simulation)
is also quartered.  Only when this scaling is verified can one be sure
that one is working in the correct asymptotic regime.

The presence of an interparticle potential adds an additional length scale
-- the range $y$ of the force -- and therefore a new dimensionless
parameter $\lambda/y$, or equivalently $c/y$.  To derive the flux-field
decomposition, we demanded that this ratio be small, but of order unity;
that is, we did not scale this parameter with the Knudsen number.
Operationally, this means that every time the size of the lattice is
doubled, the range of the force in lattice units should be kept the same.
If it is ten lattice units at one resolution, it should be ten lattice
units at all resolutions\footnote{In some sense, this means that lattice
artifacts never completely disappear, as the set of sites within a
ten-lattice-unit radius of a given site are not distributed uniformly or
isotropically.  If this is deemed problematic, it may be possible to scale
$c/y$ with $\mbox{Kn}^{-1/2}$, or in some other way such that both $c/y$
and $y/L$ vanish in the scaling limit; this has the attraction of
completely removing such problems in the scaling limit, but further
exploration of such considerations lies outside the scope of this paper.}.

While Eq.~(\ref{eq:dvc}) for $\Delta V_c$ is exact, the flux-field
decomposition of Eq.~(\ref{eq:dvcff}) is usually used to approximate
$\Delta V_c$ only to some specified order in $c/y$.  Having determined
that the RK model is just such an approximation, we are now in a
position to examine how the error incurred by this approximation scales
in the continuum limit.  Let us compare the RK model to a variant of our
model, in which we adopt the strategy of permitting the number of terms
$n_{\max}$ that we retain in the Taylor expansion for $\Delta V_c$ to
increase in the hydrodynamic limit.  We shall justify this strategy a
posteriori.  Specifically, let us take $\xi$ more terms each time the
lattice size $N$ is doubled.  If we take the system size in physical
units $L$ to be fixed, then the lattice spacing is $c\sim L/N$.  For a
dense lattice gas, the mean-free path is of order $c$, so the Knudsen
number Kn scales as $c/L$.  It follows that
\begin{equation}
n_{\max} = \xi\log_2\left(\frac{N}{N_0}\right)
         = n_0 - \xi\log_2\left(\mbox{Kn}\right),
\end{equation}
where $N_0$ and $n_0$ are constants.  If we then take the interaction
range in lattice units $y/c$ to be fixed, then the error term
$\varepsilon$ in the Taylor expansion goes as
\begin{equation}
\varepsilon
 \sim\left(\frac{c}{y}\right)^{n_{\max}}
 \sim\left(\frac{c}{y}\right)^{n_0}
     \left(\frac{y}{c}\right)^{\xi\log_2 (\mbox{\scriptsize Kn})}
 \sim\left(\frac{c}{y}\right)^{n_0}
     \mbox{Kn}^{\xi\log_2 (\frac{y}{c})}.
\end{equation}
For $\xi=1$ (adding one more term to the series at each lattice
refinement), it follows that by keeping the lattice spacing less than half
of the characteristic interaction range, the error will scale as the
Knudsen number; likewise, by making the lattice spacing less than a fourth
of the characteristic interaction range, the error will scale as the square
of the Knudsen number; and so on.  One can also increase $\xi$ to raise the
power of the Knudsen number to which the error scales.  Thus, for small
{\it but finite} values of $c/y$ (order unity in the scaling limit), the
error can be made to scale subdominantly to terms that are usually
neglected in a Chapman-Enskog expansion anyway, providing the a posteriori
justification promised above.

A potential problem with the RK model is that it does not refine the
definition of the energy in this way at each level of the scaling limit
($\xi=0$), and so the corrections that it neglects may indeed matter in
that limit.  Of course, in most situations, one will choose a particular
fixed value of $n_{\max}$; in fact, all studies of the RK model to date
have used $n_{\max}=1$, simply because it has not been previously
recognized that the RK model is only the first-order approximation to a
more exact model.  The RK model is known to exhibit certain anomalous
phenomena; for example {\it spurious currents} are known to develop near
interfaces, even if there is no bulk flow, and the surface tension is
known to be slightly anisotropic~\cite{bib:aniso}.  It is possible that
such anomalies would be eliminated by the more exact treatment of the
scaling limit advocated here, but a numerical test of this conjecture is
outside the scope of the present work.

Of course, such anomalies may be regarded as tolerable as long as one
appreciates that one is working only to lowest order of an asymptotic
series.  Indeed, because the series is asymptotic, there is no point in
being overzealous about the value of $n_{\max}$, since the series may
begin to diverge at some point.  There is at least one situation,
however, in which the observation that it is necessary to let $n_{\max}$
scale with Knudsen number may be critically important, and that is when
one is studying the scaling of (possibly divergent) quantities with
system size.  For example, Rothman and Flekk{\o}y~\cite{bib:rfl}
recently studied the scaling properties of fluctuating interfaces using
the RK model, measuring, among other things, the saturated width of the
interface as a function of system size.  Superimposed upon the usual
power-law behavior of the saturated width, they found a logarithmic
correction that resisted theoretical explanation.  We suggest that this
anomaly may be due to scaling the system size with fixed $n_{\max}$
(since the RK model effectively fixes $n_{\max}$ at unity), rather than
letting $n_{\max}$ increase linearly with system size as advocated here.
Again, it would be interesting to verify this conjecture by redoing the
numerical experiment using our model, but that is outside the scope of
the present work.

The main point of this section has been to demonstrate that the
flux-field decomposition allows one to understand in what sense the RK
model is an approximation to an exact interaction potential, why it may
be used for trial state sampling, how it might be corrected at higher
order in $c/y$, and why its scaling limit may be more subtle than
previously suspected.  We emphasize that the strategy of scaling
$n_{\max}$ with $\mbox{Kn}$ was invoked only to facilitate this
discussion of the scaling limit, and we are certainly not suggesting
that it be used in practical simulations.  If one is planning to work at
any order of $c/y$ at which $\Delta V_c$ involves nonlocal interactions,
it makes much more sense to use Eq.~(\ref{eq:dvc}) directly than to try
to deal with higher order terms of the flux-field decomposition.  For
Monte Carlo sampling of the outgoing states, it is possible that the
lowest-order local term of the flux-field decomposition could be used
for sampling, while the exact expression for $\Delta V_c$ could be used
for the acceptance criterion.

\section{Conclusions}

We have shown that the lattice gas model developed by Rothman and Keller
for immiscible-fluid hydrodynamics can be derived from an underlying
model of particle interactions.  From the enhanced understanding
provided by our observation, we elucidated the nature of the
hydrodynamic limit of the Rothman-Keller model, demonstrating that it is
more subtle than previously suspected.  Though practical simulations of
the particulate model are likely to be significantly more
compute-intensive than the original version of the Rothman-Keller model,
this work is offerred in the spirit that it is always useful to know the
exact model corresponding to any given approximation.  We hope that this
work helps to provide some theoretical basis for these models' success,
and perhaps for their ultimate improvement.

\section*{Acknowlegements}

BMB would like to thank the International Centre for Theoretical Physics
(ICTP) for their hospitality during a portion of this work.  BMB was
supported in part by the United States Air Force Office of Scientific
Research under grant number F49620-95-1-0285.  The collaboration of BMB
and PVC was facilitated by NATO grant number CRG 950356.

\newpage
\appendix

\section{Derivation of Flux-Field Decomposition}
\label{app:a}

To derive Eq.~(\ref{eq:dvcff}) for $\Delta V_c$, we begin with the Taylor
expansion of a function of the magnitude of a displaced vector,
\begin{equation}
\phi(|\bfy+\bfc|) =
 \sum_{n=0}^\infty
 \frac{\epsilon^n}{n!}
 \sum_{m=\lceil n/2 \rceil}^n
 \frac{n!\phi_m(y)}{2^{n-m}(2m-n)!(n-m)!}
 \left(\bfy\cdot\bfc\right)^{2m-n}
 \left(\bfc\cdot\bfc\right)^{n-m},
\end{equation}
where $\epsilon$ has been introduced as an expansion parameter to keep
track of the order in $c/y$ (it is numerically equal to one), $y\equiv
|\bfy|$, and we have defined the following functions related to the
derivatives of $\phi (y)$:
\begin{equation}
\phi_m(y)\equiv
 \left(\frac{1}{y}\frac{d}{dy}\right)^m
 \phi(y).
\end{equation}
We let $\bfc\rightarrow \bfc_j-\bfc_i$ and use the binomial theorem to
write
\begin{eqnarray}
\left(\bfy\cdot\bfc\right)^{2m-n}
 &=&
 \left(\bfy\cdot\bfc_j - \bfy\cdot\bfc_i\right)^{2m-n}\nonumber\\
 &=&
 \sum^{2m-n}_{l=0}
 \frac{(2m-n)!}{l! (2m-n-l)!}
 \left(\bfy\cdot\bfc_j\right)^{2m-n-l}
 \left(-\bfy\cdot\bfc_i\right)^l,
\end{eqnarray}
and
\begin{eqnarray}
\left(\bfc\cdot\bfc\right)^{n-m}
 &=&
 \left(|\bfc_j|^2 + |\bfc_i|^2 - 
       2 \bfc_j\cdot\bfc_i\right)^{n-m}\nonumber\\
 &=&
 \sum^{n-m}_{k=0}
 \sum^{n-m-k}_{p=0}
 \frac{(n-m)!|\bfc_i|^{2k} |\bfc_j|^{2p}}{k! p! (n-m-k-p)!}
 \left(-2 \bfc_i\cdot\bfc_j\right)^{n-m-k-p}.
\end{eqnarray}
Inserting these into Eq.~(\ref{eq:dvc}) for $\Delta V_c$, we get
\begin{eqnarray}
\lefteqn{\Delta V_c  =}\nonumber\\
 & &
 \smallfrac{1}{2}
 \sum_{\bfx,\bfy}\sum_{i,j}
 q_i'(\bfx)q_j'(\bfx+\bfy)
 \sum_{n=1}^\infty
 \frac{\epsilon^n}{n!}
 \sum_{m=\lceil n/2 \rceil}^n
 \nonumber\\
 & &
 \indentspace
 \sum_{l=0}^{2m-n}
 \sum_{k=0}^{n-m}
 \sum_{p=0}^{n-m-k}
 \frac{(-1)^{n-m+l-k-p} n!}{2^{k+p}l!k!p!(2m-n-l)!(n-m-k-p)!}
 \nonumber\\
 & &
 \indentspace\indentspace\phi_m(y)
 \left(\bfy\cdot\bfc_j\right)^{2m-n-l}
 \left(\bfy\cdot\bfc_i\right)^l
 |\bfc_i|^{2k} |\bfc_j|^{2p}
 \left(\bfc_i\cdot\bfc_j\right)^{n-m-k-p}.
\end{eqnarray}
Eliminating $p$ in favor of the new summation index $r\equiv n-m+l+k-p$,
and reordering the summations, this becomes
\begin{eqnarray}
\lefteqn{\Delta V_c  =}\nonumber\\
 & &
 \smallfrac{1}{2}
 \sum_{n=1}^\infty
 \frac{\epsilon^n}{n!}
 \sum_{r=0}^n
 \sum_{\bfx,\bfy}\sum_{i,j}
 q_i'(\bfx)q_j'(\bfx+\bfy)
 \sum_{m=\lceil n/2 \rceil}^n
 \sum_{l=0}^{2m-n}\,
 \sum_{k=\max (0,m-n+r-l)}^{\min (n-m,\lfloor(r-l)/2\rfloor)}
 \nonumber\\
 & &
 \indentspace
 \frac{(-1)^r n!\phi_m(y) |\bfc_i|^{2k} |\bfc_j|^{2n-2r-2m+2l+2k}}
      {2^{n-r-m+l+2k}l!k!(n-r-m+l+k)!(2m-n-l)!(r-l-2k)!}
 \nonumber\\
 & &
 \indentspace
 \left(\bfy\cdot\bfc_j\right)^{2m-n-l}
 \left(\bfy\cdot\bfc_i\right)^l
 \left(\bfc_i\cdot\bfc_j\right)^{r-l-2k},
\end{eqnarray}
where we have adopted the convention that a sum is zero if its upper
limit is less than its lower limit.  By reinterpreting dot products
raised to the $s$ power as the $s$-fold inner product of two $s$-fold
outer products, we can rewrite this as follows
\begin{eqnarray}
\Delta V_c 
 &=&
 \smallfrac{1}{2}
 \sum_{n=1}^\infty
 \frac{\epsilon^n}{n!}
 \sum_{r=0}^n
 (-1)^r \lchoose{n}{r}
 \sum_{\bfx,\bfy}\sum_{i,j}
 q_i'(\bfx)q_j'(\bfx+\bfy)\nonumber\\
 & &
 \indentspace
 \left[
 \left(\bigotimes^r\bfc_i\right)
 \bigodot^r
  \calk_n (\bfy)
 \bigodot^{n-r}
 \left(\bigotimes^{n-r}\bfc_j\right)
 \right],
 \label{eq:dvaa}
\end{eqnarray}
where $\bigotimes^r$ denotes an $r$-fold outer product and $\bigodot^r$
denotes an $r$-fold inner product, and where we have defined the kernel
\begin{eqnarray}
\calk_n (\bfy)
 & \equiv &
 \sum_{m=\lceil n/2 \rceil}^n
 \sum_{l=0}^{2m-n}\,
 \sum_{k=\max (0,m-n+r-l)}^{\min (n-m,\lfloor(r-l)/2\rfloor)}
 \nonumber\\
 & &
 \frac{r! (n-r)!\phi_m(y)}
      {2^{n-r-m+l+2k}l!k!(n-r-m+l+k)!(2m-n-l)!(r-l-2k)!}
 \nonumber\\
 & &
 \indentspace
 \left[
  \left(\bigotimes^{2m-n}\bfy\right)
  \otimes
  \left(\bigotimes^{n-m}\bfone\right)
 \right]
 \nonumber\\
 & = &
 \sum_{m=\lceil n/2 \rceil}^n
 \frac{n!\phi_m(y)}{2^{n-m}(2m-n)!(n-m)!}
 \left[
  \left(\bigotimes^{2m-n}\bfy\right)
  \otimes
  \left(\bigotimes^{n-m}\bfone\right)
 \right].
\label{eq:kernela}
\end{eqnarray}
In the very last step above we performed the sums over $k$ and $l$.

The expression for the potential energy, Eq.~(\ref{eq:dvaa}), has a
remarkable symmetry, inherited from Eq.~(\ref{eq:dv}).  By making the
substitutions
\begin{eqnarray}
 r &\leftarrow& n-r\nonumber\\
 i &\leftarrow& j\nonumber\\
 j &\leftarrow& i\nonumber\\
 \bfx &\leftarrow&\bfx+\bfy\nonumber\\
 \bfy &\leftarrow&-\bfy
 \label{eq:symsa}
\end{eqnarray}
and noting that $\calk_n(-\bfy)=(-1)^n\calk_n(\bfy)$, we can see
that the $r$th term of Eq.~(\ref{eq:dvaa}) is equal to the $(n-r)$th
term.  It also follows that the kernel $\calk_n$ can be chosen to be
completely symmetric under interchange of any two of its $n$ indices.
If we notice that the combinatorial factor
\begin{equation}
\frac{n!}{2^{n-m} (2m-n)!}
\end{equation}
is precisely equal to the number of distinct ways to assign $n$ indices
to the tensor $(\bigotimes^{2m-n}\bfy)\otimes (\bigotimes^{n-m}
\bfone)$, and recalling that only the symmetric part of $\calk_n$
matters, we can rewrite the kernel, Eq.~(\ref{eq:kernela}), in the
remarkably compact form of Eq.~(\ref{eq:kernel}).  The simplicity of
this result suggests that there may be an easier way to derive it.

If we now introduce the completely symmetric $r$th-rank outgoing {\it
  tensor fluxes}, as defined in Eq.~(\ref{eq:fluxes}), then
Eq.~(\ref{eq:dvaa}) may be written as
\begin{eqnarray}
\lefteqn{\Delta V_c  =}\nonumber\\
 & &
 \smallfrac{1}{2}
 \sum_{\bfx,\bfy}
 \sum_{n=1}^\infty
 \frac{\epsilon^n}{n!}
 \sum_{r=0}^n
 (-1)^r\lchoose{n}{r}
 \calj_r' (\bfx)
 \bigodot^r
 \calk_n (\bfy)
 \bigodot^{n-r}
 \calj_{n-r}' (\bfx+\bfy).
\end{eqnarray}
Because of the symmetry, Eq.~(\ref{eq:symsa}), we can remove the factor
of $1/2$ and sum $r$ from $\lceil n/2\rceil$ to $n$, instead of from $0$
to $n$.  The exception to this arises when $n$ is even and $r=n/2$; in
that case, the factor of $1/2$ must be retained.  We accommodate this
case by dividing by $1+\delta_{n,2r}$, where $\delta$ is the Kronecker
delta.  If we then define the $r$th-rank {\it tensor fields}, as in
Eq.~(\ref{eq:fields}), the expression of Eq.~(\ref{eq:dvcff}) for the
potential energy change follows immediately.

As noted in the text, the primes on the fluxes and fields indicate that
these are evaluated using the {\it outgoing} (post-collision) states.
Thus, at each order $n$, Eq.~(\ref{eq:dvcff}) expresses the change in
potential energy due to the propagation step as the sum of $n$ terms,
the $r$th of which is an $r$-fold inner product of an $r$th-rank
outgoing tensor flux, Eq.~(\ref{eq:fluxes}), with an $r$th-rank tensor
field, Eq.~(\ref{eq:fields}).

Note that we can rearrange the order of summation of $n$ and $r$ to
write Eq.~(\ref{eq:dvcff}) in the following alternative format
\begin{equation}
\Delta V_c 
 =
 \sum_\bfx
 \sum_{r=1}^\infty
 \epsilon^r
 \calj_r' (\bfx)
 \bigodot^r
 \left(
  \sum_{n=0}^r
  \epsilon^n
  \cale_{n+r,r}' (\bfx)
 \right).
\end{equation}
This makes it clear that at each order in $r$ a new $r$th-rank tensor
flux must be introduced, and the corresponding $r$th-rank field is the
sum of $r$ terms.  The disadvantage of writing the result in this way is
that each term of the outer sum over $r$ contains terms of differing
order in $\epsilon$.

Finally, we note in passing that tensor fluxes and fields have been
considered in the context of a LGA model of
microemulsions~\cite{bib:bce}.  While that model did indeed derive its
update rule from considerations of single-particle interactions, it did
not employ the method used here.  In particular, where $\bfc_j-\bfc_i$
appears in Eq.~(\ref{eq:dv}), that study took only $-\bfc_i$, and the
expansion was not carried out to all orders.  On the other hand, that
work also included vector-valued charge attributes, in order to properly
model the orientation of the surfactant molecules.  It is likely that a
more exact formulation, analogous to our Eq.~(\ref{eq:dv}), exists for
this microemulsion model, and may well connect this work with the static
lattice models of microemulsions due to Matsen and
Sullivan~\cite{bib:eque}.


\begin{thebibliography}{999}
  \parindent=.6em 
\bibitem{bib:rk} D.H. Rothman and J.M. Keller, {\it Phys. Rev. Lett.}
  {\bf 56}, 889 (1988).
\bibitem{bib:rz} D.H. Rothman and S. Zaleski, {\it Lattice-Gas
    Automata: Simple Models of Complex Hydrodynamics}, (Cambridge
  University Press, 1997).
\bibitem{bib:fhp} U. Frisch, B. Hasslacher, and Y. Pomeau, {\it
    Phys. Rev. Lett.} {\bf 56} 1505 (1986).
\bibitem{bib:swolf} S. Wolfram, {\it J. Stat. Phys.} {\bf 45}, 471
  (1986).
\bibitem{bib:chen} H. Chen, S. Chen, G.D. Doolen, Y.C. Lee, {\it
    Phys. Rev. A} {\bf 40}, 2850-2853 (1989).
\bibitem{bib:cl} C.K. Chan and N.Y. Liang, {\it Europhys. Lett.} {\bf
    13}, 495-500 (1990).
\bibitem{bib:fchc} U. Frisch et al., {\it Complex Syst.} {\bf 1}, 648
  (1987).
\bibitem{bib:ll} L. L. Landau and E. M. Lifschitz  {\it Fluid
    Mechanics}, (Pergamon, New York, 1982), p. 24.
\bibitem{bib:aniso} C. Adler, D. D'Humi\`{e}res, D.H. Rothman, {\it
    J. Phys. I France} {\bf 4}, 29-46 (1994).
\bibitem{bib:rfl} E. G. Flekk{\o}y and D. H. Rothman, {\it Phys. Rev. E}
  {\bf 53}, 1622 (1996); {\it Phys. Rev. Lett.} {\bf 75}, 260 (1995).
\bibitem{bib:bce} B. Boghosian, P. Coveney, and A. Emerton, {\it
    Proc. Roy. Soc. A} {\bf 452} (1996) 1221.
\bibitem{bib:eque} M.W. Matsen, D.E. Sullivan, {\it Phys. Rev. A} {\bf
    46}, 1985-1991 (1992); {\it Phys. Rev. E} {\bf 51}, 548-557 (1995).
\end{thebibliography}
\end{document}